# SPACE CHARGE EFFECTS IN CYCLOTRON GAS STOPPER

Yuri K. Batygin

Los Alamos National Laboratory, Los Alamos, NM 87545

**Abstract**

The cyclotron gas stopper is a newly proposed device to stop energetic rare isotope ions from projectile fragmentation reactions in a helium-filled chamber [1, 2]. The radioactive ions are slowed down by collisions with a buffer gas inside a cyclotron-type magnet and are extracted via interactions with a Radio Frequency (RF) field applied to a sequence of concentric electrodes (RF carpet). The present study focuses on a detailed understanding of space charge effects in the ion extraction region. The space charge is generated by the ionized helium gas created by the stopping of the ions and eventually limits the beam rate. Particle-in-cell simulations of a two-component (electron-helium) plasma interacting via Coulomb forces were performed in the space charge field created by the stopping beam.

Email: batygin@lanl.gov

## 1. Introduction

The cyclotron gas stopper is a device for the deceleration of radioactive ions created by the projectile fragmentation (see Fig. 1). Fast ions (~100 MeV/u) are injected into a helium-filled chamber inside a vertical magnetic field where they immediately enter a solid degrader so that they can be captured by the magnetic field. The fast ions lose the remainder of their kinetic energy in collisions with the helium buffer gas. This process ionizes the helium atoms. An electric field parallel to the magnetic field is used to remove electrons and move positively charged ions to the RF-carpet. At high incoming particle rates, the amount of ionization becomes so large that the ions cannot be completely removed. As a result, a neutralized plasma accumulates in the center of the stopping chamber and additional fast ions are not or are only slowly extracted because they come to rest in the plasma-shielded region. This present work analyzed the overall process of charge migration. In contrast with previous studies [1-4], we use self-consistent particle-in-cell model for space-charge effects, where both helium and electron components move in the field defined by that motion. All other effects except Coulomb interaction between particles are neglected.

## 2. Numerical Method

Present simulations are based on a preceding detailed numerical study of rare isotope production, transport, and stopping in a gas-filled magnetic field. The program LISE++ [5] was used to calculate the transmission, yields, and ion-optical properties of the projectile fragment beam. A C++ version of the ATIMA code [6] was used to calculate the energy lost by the incoming beam in the solid degrader. The Stopping and Range Tables from the SRIM package [7] were used to calculate the energy loss in the helium gas. The CycStop code [8] combined this input to calculate the fast ion stopping distribution, the losses, and the spatial deposition of energy in the helium. The energy distributions were the input for the space charge phenomena in the present work (see Fig. 2).

The calculation of space charge effects in $e^-/He^+$ plasma were performed with a modified version of the BEAMPATH code [9]. The simulations were performed by simultaneous tracking of $He^+$ and electrons in the field created by their own space charge forces, $\vec{E}_{sc}$, and applied external electric field, $\vec{E}_o$, with a velocity $\vec{v}$ given by

$$\vec{v} = k\,(\vec{E}_o + \vec{E}_{sc}), \tag{1}$$

where $k$ is the mobility of ion or electron. The mobility of $He^+$ ions at room temperature and pressure of $P$ = 200 mbar was taken as $k_{He+}$ = 81 $cm^2/(Vs)$ and is inversely proportional to gas pressure. The mobility of electrons is typically three orders of magnitude higher than that of $He^+$ under the same conditions. However, direct simulation of $e^-/ He^+$ plasma with such large differences in ion mobilities is problematic due to the dramatic difference in the time scales of the dynamics of electrons and $He^+$. Therefore, in the present work the mobility of the electrons was artificially reduced by a factor or ten to $10^4$ $cm^2/(Vs)$ to simultaneously follow the motion of the ions and electrons in the same calculation.

Radioactive ions were injected into the gas at a rate of $dN/dt_{in}$. Each incoming radioactive ion created a cloud of $He^+$ and electron charge at the rate $dQ/dN$, each. The product of these two rates gives the rate of creation of $He^+$ and electrons in the gas-filled chamber

$$J = \frac{dQ}{dt} = \frac{dN}{dt}\frac{dQ}{dN}\,. \tag{2}$$

The simulation time step, $\Delta t = \Delta T/N_{step}$, was defined as a ratio of the simulation period, $\Delta T$, and the number of integration steps, $N_{step}$. At each integration step the entire spatial distribution of $e^-/He^+$ particles was injected into the system such that the incoming charge of each component was

$$\Delta q = \frac{dQ}{dt}\Delta t. \tag{3}$$

This distribution is added to that remaining from the previously injected particles (see Fig. 4). The electrons and positive ions move in opposite directions in the net field created by applied collection field and by their space charge.

Space charge field is calculated via solution of Poisson's equation in two-dimensional cylindrical coordinates:

$$\frac{1}{r}\frac{\partial}{\partial r}\left(r\frac{\partial U}{\partial r}\right) + \frac{\partial^2 U}{\partial z^2} = -\frac{\rho(r,z)}{\varepsilon_o}, \tag{4}$$

where $\rho(r, z)$ is the space charge density, $U(r,z)$ is the space charge potential, with Dirichlet boundary conditions at the metallic surfaces of a tube of radius $a$, and Neumann condition at the axis (see Fig. 3):

$$U(a,z) = 0, \qquad U(r,0) = U(r,L) = 0, \qquad \frac{\partial U}{\partial r}(0,z) = 0. \tag{5}$$

In Eq. (4), $\rho(r,z)$ is the total instantaneous value of charge of electrons and ions presented in the system. While at every time step, the new fraction of $e^-/He^+$ is injected into the system, some fraction of particles is removed through the extraction region and through the walls of the gas cell. The Poisson's equation (4) is substituted by finite-difference analog on the grid $N_r \times N_z$:

$$U_{k,j+1}\left(1 + \frac{1}{2(j-1)}\right) - 2U_{k,j}\left(1 + \frac{h_r^2}{h_z^2}\right) + U_{k,j-1}\left(1 - \frac{1}{2(j-1)}\right)$$

$$+ U_{k+1,j}\left(\frac{h_r}{h_z}\right)^2 + U_{k-1,j}\left(\frac{h_r}{h_z}\right)^2 = -\frac{\rho_{k,j}}{\varepsilon_o}h_r^2. \tag{6}$$

Poisson’s equation is solved via Fourier transformation in z-direction and solution of three-diagonal matrix equation in radial direction. Accuracy of calculations is controlled by calculation of error of the Gauss theorem

$$\vartheta = 1 - \left| \frac{\int_S \vec{E}\, d\vec{S}}{\frac{1}{\varepsilon_o}\int_V \rho\, dV} \right| , \qquad (7)$$

which is usually of the order of $10^{-3}$ (see Fig. 5). Table 1 contains numerical parameter used in simulation.

## 3. Analytical Treatment of Space Charge Problem

Particle motion in considered system is performed mostly in longitudinal direction while radial motion is not significant. Self-consistent problem for one-dimensional particle distribution with space charge field $E_z$, and external field $E_o$ can be treated analytically. We assume that particles are injected into the system from the edge of neutralized plasma region. The problem includes simultaneous solution of Poisson's equation for unknown space charge field and equation for particles motion:

$$\begin{cases} \dfrac{dE_z}{dz} = \dfrac{\rho}{\varepsilon_0} \\ v_z = k\,(E_o + E_z) \;. \end{cases} \qquad (8)$$

Current density, $j = \rho v_z$, is given by

$$j = \frac{J}{\pi R^2} \;, \qquad (9)$$

where $R$ is the radius of injected charge area. Therefore, space charge density is expressed through injected charge parameters as

$$\rho = \frac{J}{\pi R^2 v_z}. \qquad (10)$$

Substitution of Eq. (10) into Poisson's equation (8) gives for unknown space charge field:

$$\frac{dE_z}{dz} = \frac{J}{\varepsilon_o \pi R^2 k\,(E_o + E_z)}, \qquad \text{or} \qquad E_z dE_z + E_o dE_z = \frac{J}{\varepsilon_o \pi R^2 k} dz \;. \tag{11}$$

Integration of Eq. (11) results in quadratic equation for unknown space charge field $E_z$:

$$\frac{E_z^2}{2} + E_o E_z - \frac{J}{\varepsilon_o \pi R^2 k} \Delta z = 0 \,. \tag{12}$$

Solution of Eq. (12) is

$$E_z + E_o = \sqrt{E_o^2 + 2\frac{J}{\varepsilon_o \pi R^2 k} \Delta z} \;. \tag{13}$$

Table 2 illustrates total field, $E_z + E_o$, Eq. (13), at $z = 0$, as a function of outcoming charge rate. Comparison indicates good agreement between numerical and analytical values.

## 4. Simulation Results

Fig. 4 illustrates the dynamics of $e^-/He^+$ cloud formation for the injection of $^{79}$Br ions into the system with an axial electric field of $E_o$ = -10 V/cm for an incoming particle rate of $dN/dt_{in} = 10^5$ particles per second (pps) and energy of $E = 1.69\cdot10^9$ eV. The number of $e^-/He^+$ pairs per stopped ion is $N = E / E_i = 4.23\cdot10^7$ where $E_i$ = 40 eV is the effective ionization potential in helium gas. Each incoming beam particle creates $e^-/He^+$ pairs equal to $dQ/dN = eN = 6.76\cdot10^{-12}$ C/part so that the increase in charge due to $e^-/He^+$ pairs in the system is $dQ_{in}/dt = 6.76\cdot10^{-7}$ C/s. As illustrated in Fig. 4, shortly after starting the injection of beam, the electrons quickly travel to the anode wall of gas cell at $z = L = 10$ cm (the right edge in this view). After $t = 5\cdot10^{-3}$ sec, the system has relaxed to steady-state condition for this beam rate. Under these conditions the outgoing particle rate is equal to incoming rate $dQ_{out}/dt = dQ_{in}/dt$. The charged particles are considered to be extracted from the gas if they reach the walls (anode or cathode used to apply the drift field). Conditions for extraction of helium ions might be affected by an additional focusing provided by RF carpet.

At significantly higher incoming beam rates, the fraction of outgoing ions becomes lower than the creation rate of ionized gas. This results from the formation of charged neutralized area in the regions of the chamber with the highest ionization densities. The present calculations were used for study conditions of formation of charged neutralized area. The actual spiral distribution of ionized buffer gas was replaced by a uniform disk of 75 cm radius, with an axial ionization distribution centered at $z$ = 5 cm and $\sigma_z$ = 1.5 cm. Both the actual and simple distributions give a similar results under the same values of incoming charge rate.

Fig. 6 illustrates the flux of electrons through $z$ = 0 and that of $He+$ through $z$ = 10 cm at different values of incoming charge rate under pressure of $P$ =100 mbar and applied external field of $E_o$ = -10 V/cm. Because of high mobility, electrons quickly reach the wall and their outcome flux almost immediately becomes equal to the input one (in the considered time scale), while helium ions slowly propagate and reach the wall significantly later. Starting with the value of $dQ_{in}/dt = 3.38 \cdot 10^{-5}$ C/s, injected charge cannot be extracted completely, and is accumulated in the central region of the chamber. Figs. 7 and 8 illustrate the dependence of extraction efficiency of $e^-/He^+$ ions as a function of the incoming charge rate for different values of buffer gas pressure and applied electric field. The space charge forces were found to significantly limit the extraction efficiency when the incoming rate exceeded $dQ/dt_{in} = 10^{-5}$ C/s.

## 5. Conclusion

Space charge effects in newly proposed cyclotron gas stopper were studied using particle-in-cell method. Two-component electron/helium plasma creates neutralized area in the central part of the system, which results in limitation of extraction of electrons and helium ions. This limits the extraction of radioactive ions as well because some of the beam stops in the neutralized area and cannot leave the region. Quantitative estimations of incoming charge rate for creation of neutralized area are obtained as a function of gas pressure and applied electric field.

**Acknolegements**

The author is indebted to G.Bollen, C.Campbell, F.Marti, D.J.Morrissey, G.Pang, and S.Schwarz for formulation of the problem, for numerous useful discussions, and for providing critical comments.

Table 1. Simulation parameters

| | |
|---|---|
| Energy per stopped ion | $1.69\cdot10^9$ eV |
| Helium mobility coefficient, $k_{He+}$ | 81 $cm^2/(Vs)$ |
| Electron "mobility coefficient", $k_e$ | $10^4$ $cm^2/(Vs)$ |
| Number of macroparticles per filling | $2\cdot10^3$ |
| Number of integration steps | $5\cdot10^3$ |
| Total number of macroparticles | $10^7$ |
| Box for Poisson's equation $R$ x $Z$ | 150 cm x 10 cm |
| Mesh $N_R$ x $N_Z$ | 1024 x 1024 |

Table 2. Total electric field at $z = 0$ for $E_o = -5$ V/cm, $P = 200$ mbar, $k_{He+} = 81$ $cm^2/(V\ sec)$

| $\frac{dQ_{in}}{dt}$ $(\frac{C}{\sec})$ | $\frac{dQ_{out}/dt}{dQ_{in}/dt}$ | $J = \frac{dQ_{out}}{dt}$ $(\frac{C}{\sec})$ | $\Delta z$ (cm) | $E_{numeric}$ $(\frac{kV}{cm})$ | $E_{analytical}$ $(\frac{kV}{cm})$ |
|---|---|---|---|---|---|
| $1.69\cdot10^{-6}$ | 1 | $1.69\cdot10^{-6}$ | 4.5 | 0.012 | 0.012 |
| $6.76\cdot10^{-6}$ | 0.6 | $4.06\cdot10^{-6}$ | 4 | 0.016 | 0.017 |
| $3.38\cdot10^{-5}$ | 0.25 | $8.45\cdot10^{-6}$ | 3.5 | 0.021 | 0.022 |
| $1.35\cdot10^{-4}$ | 0.1 | $1.35\cdot10^{-5}$ | 3 | 0.027 | 0.026 |

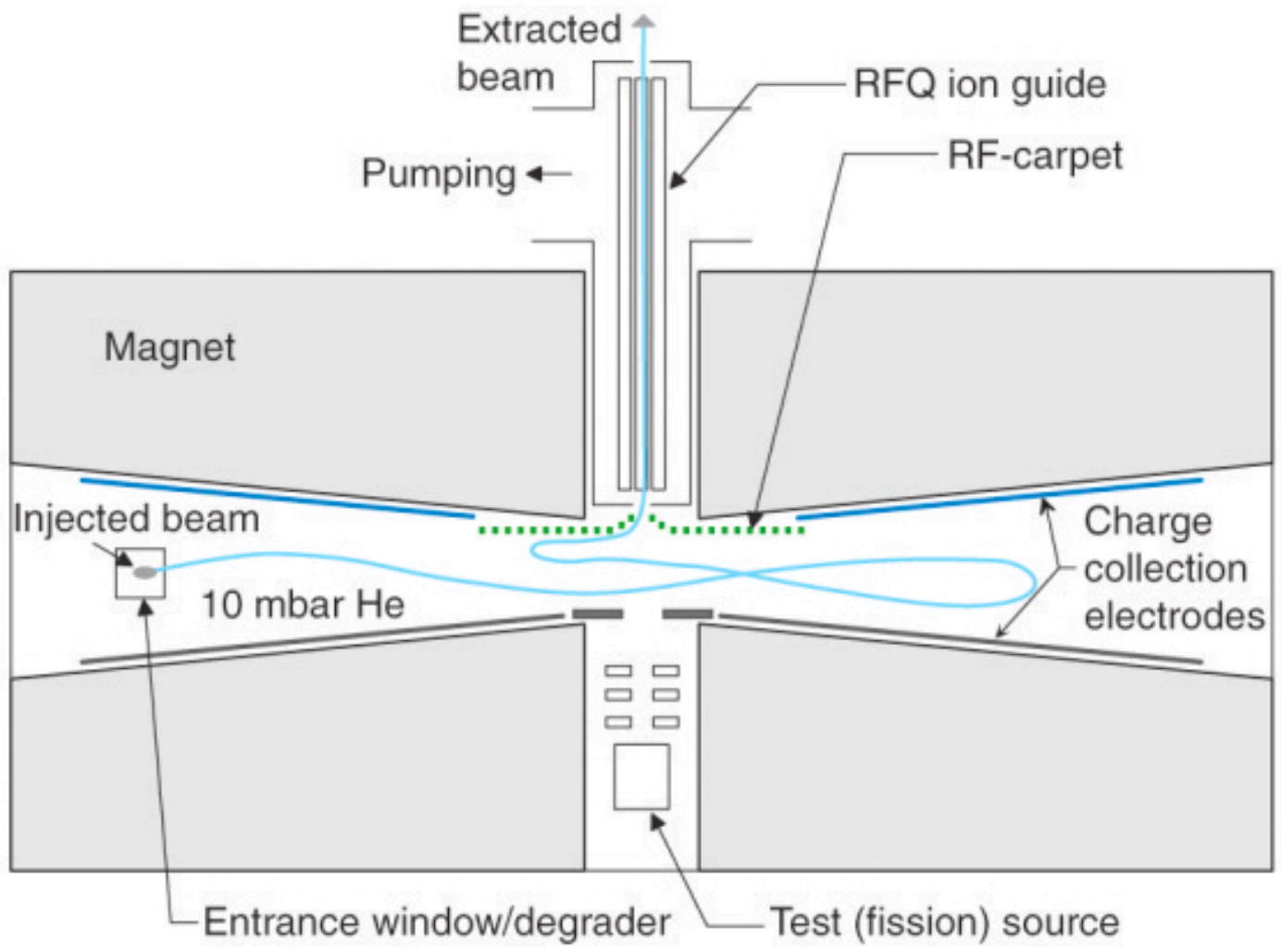

Fig. 1. Schematic layout of cyclotron gas stopper [2]. The fast projectile fragments are incident horizontally at the left and after stopping are moved to the center for axial extraction.

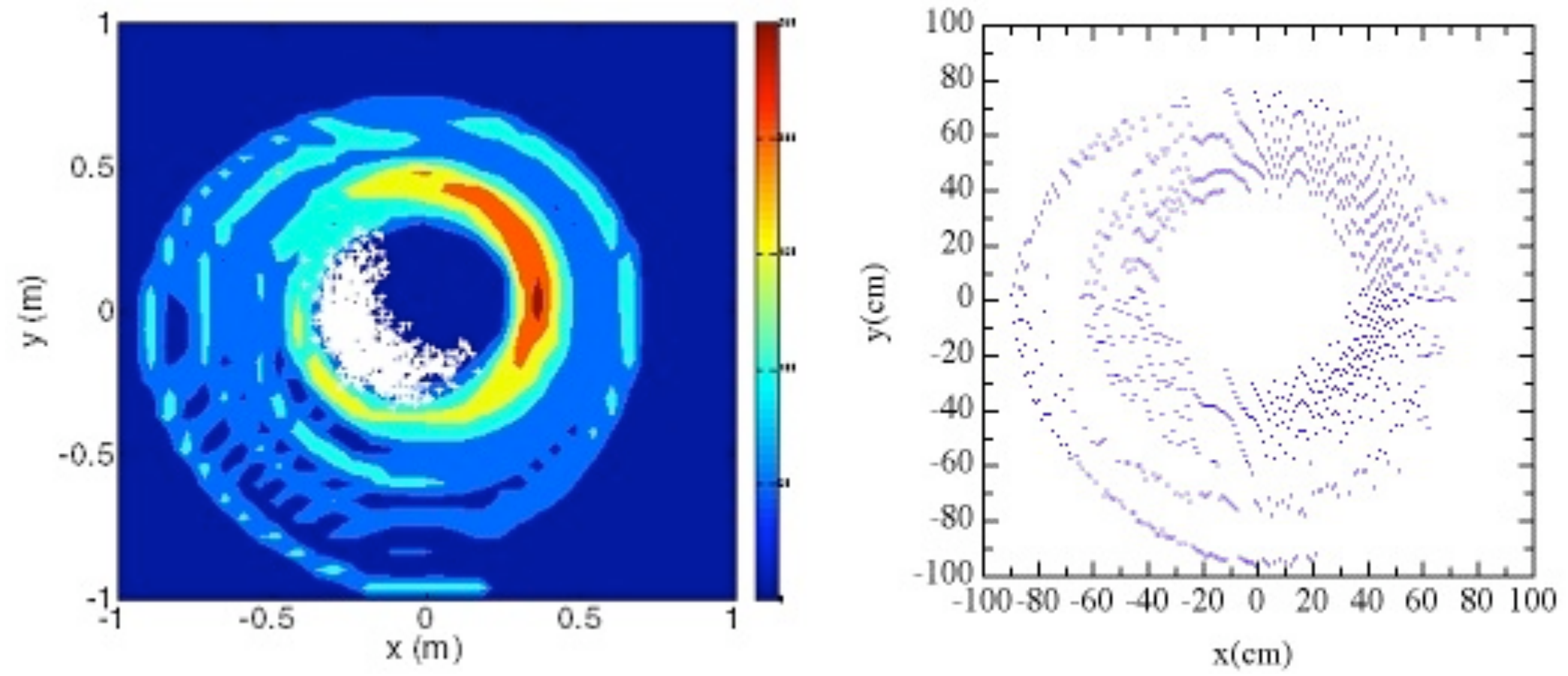

Fig. 2: (Left) Top view of the energy loss density in color and the positions of the stopped ions in white, and (right) the distribution of $e^-/He^+$ ion-pairs created by the stopping of a $^{79}$Br beam, calculated by C.Campbell using CycStop code [6].

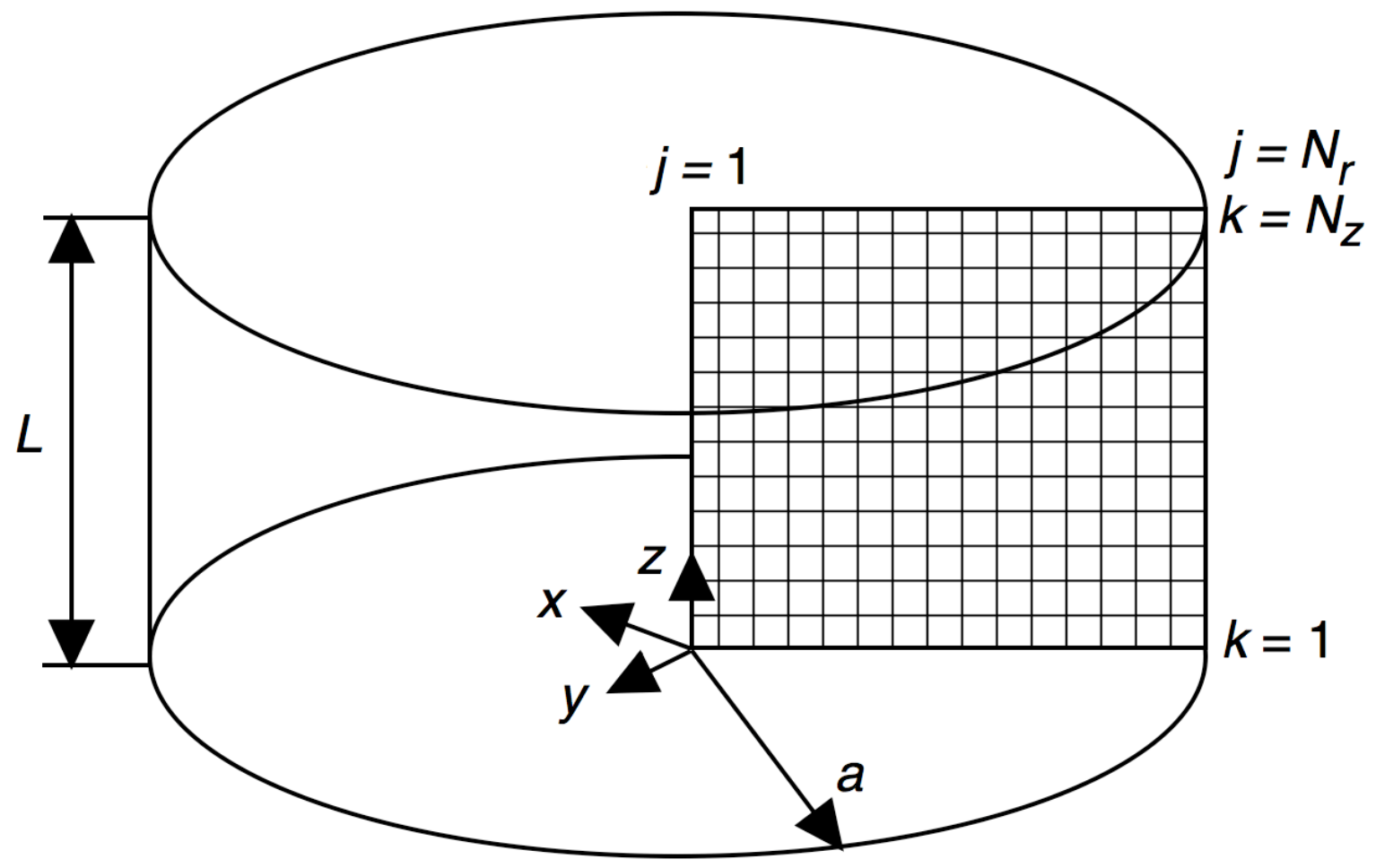

Fig. 3. Layout of numerical setup.

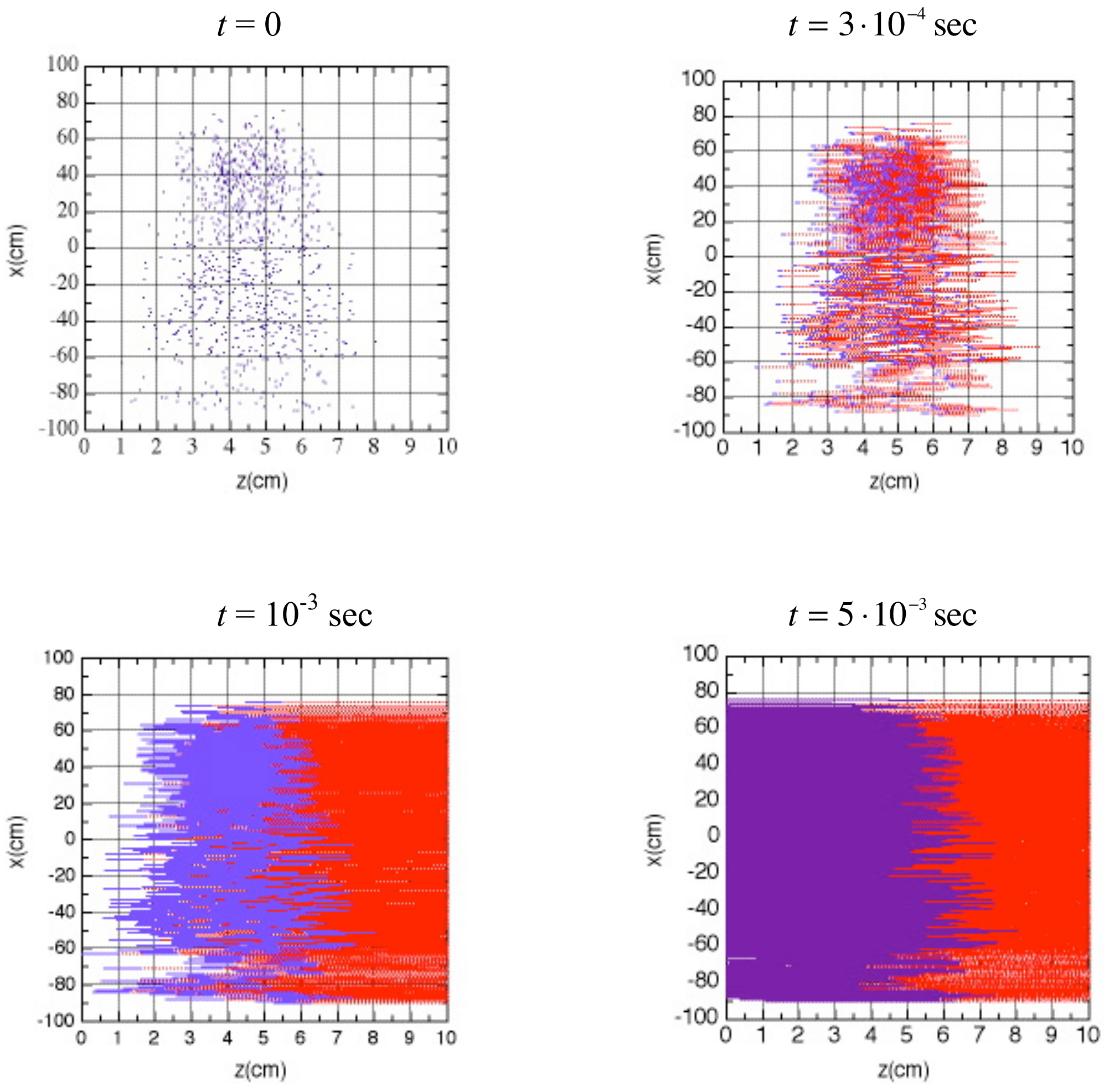

Fig. 4. Time sequence of the motion of $He^{+}$ ions (blue) and $e^{-}$(red) created by a $^{79}Br$ beam at $dN/dt_{in}$ =$10^{5}$ pps. The applied (horizontal) electric field was $E_o$ = -10 V/cm.

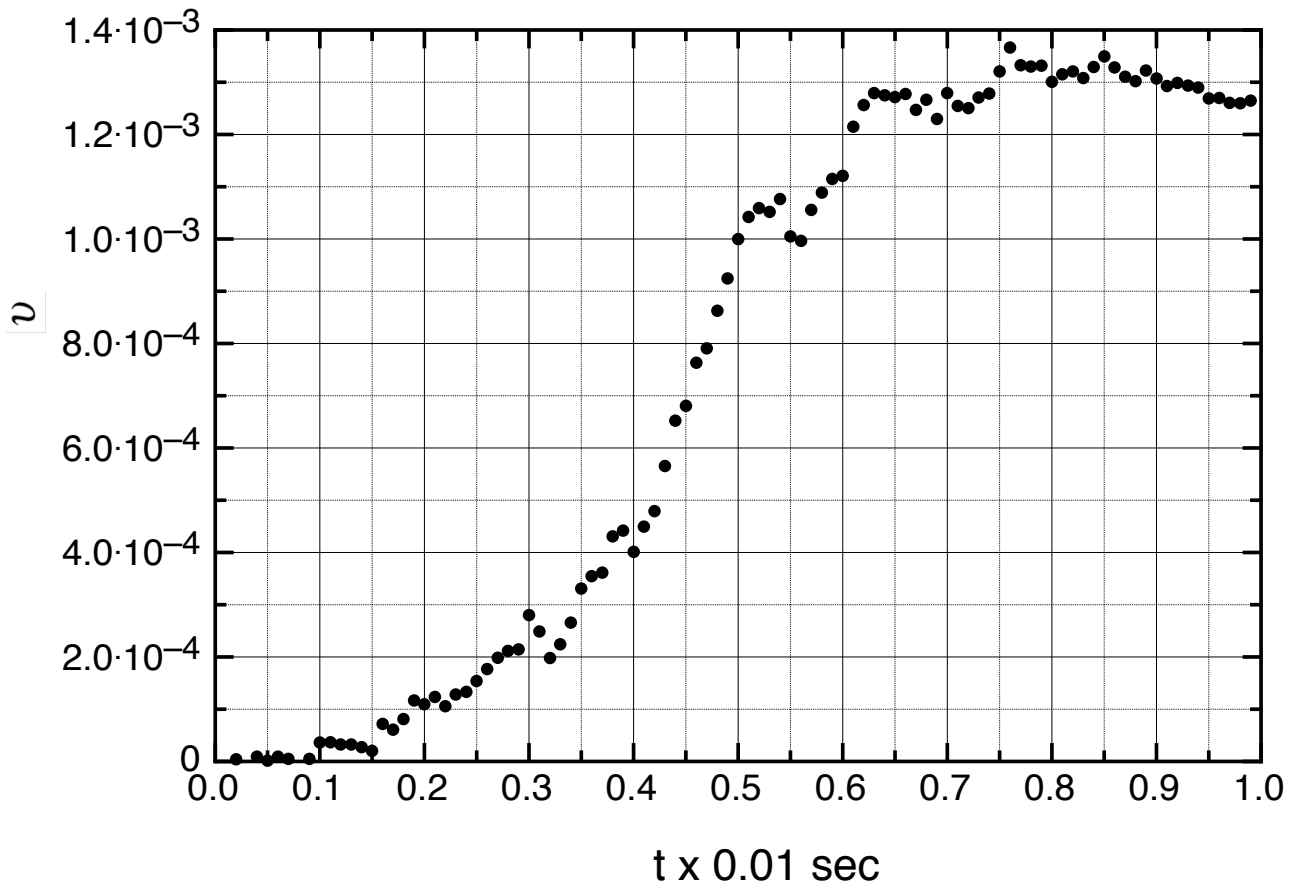

Fig. 5. Error in Gauss theorem as a function of time.

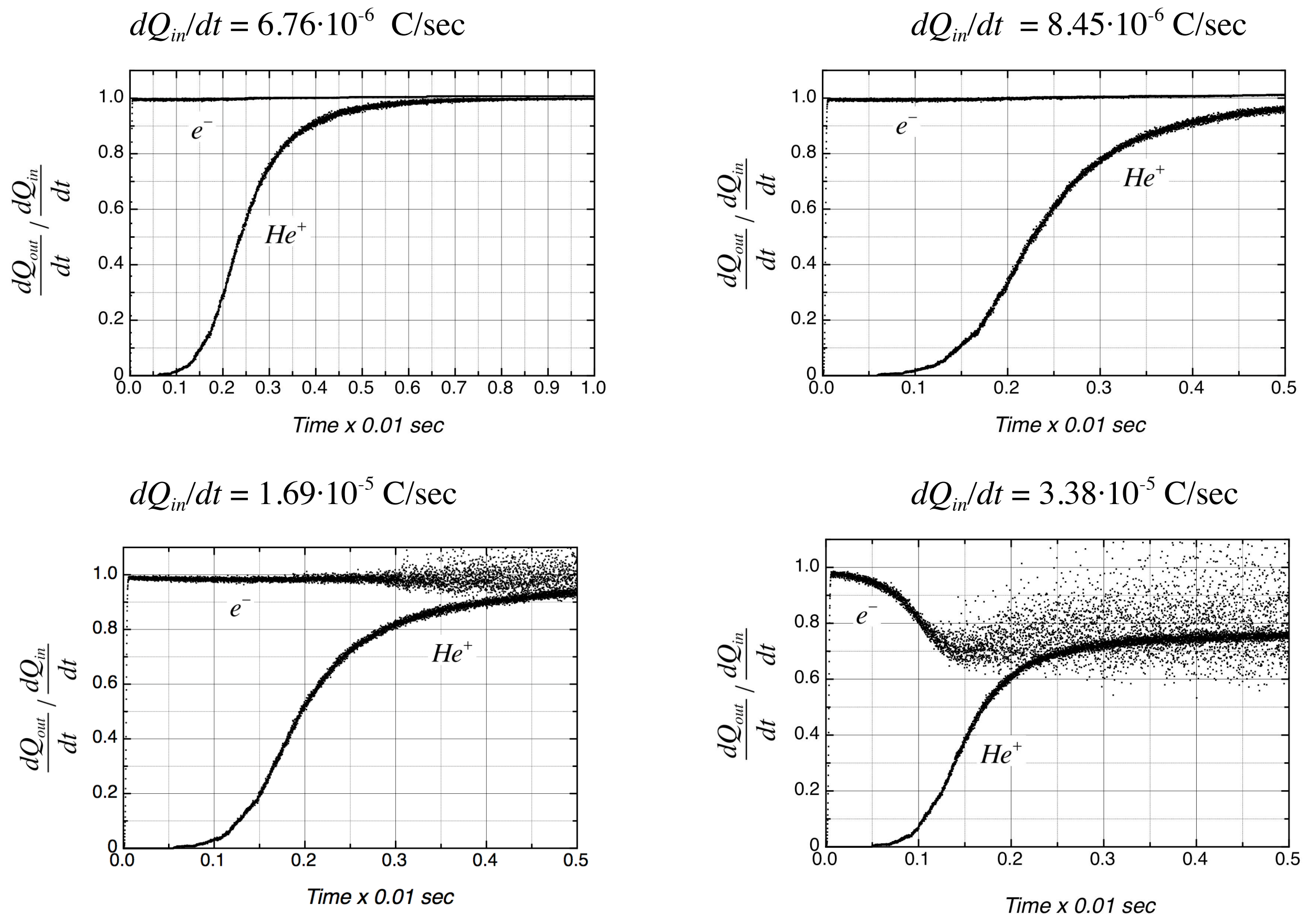

Fig. 6. Flux of electrons through $z$ = 0 and that of $He+$ through $z$ = 10 cm at different values of incoming charge rate, $P$ =100 mbar, $E_o$ = -10 V/cm.

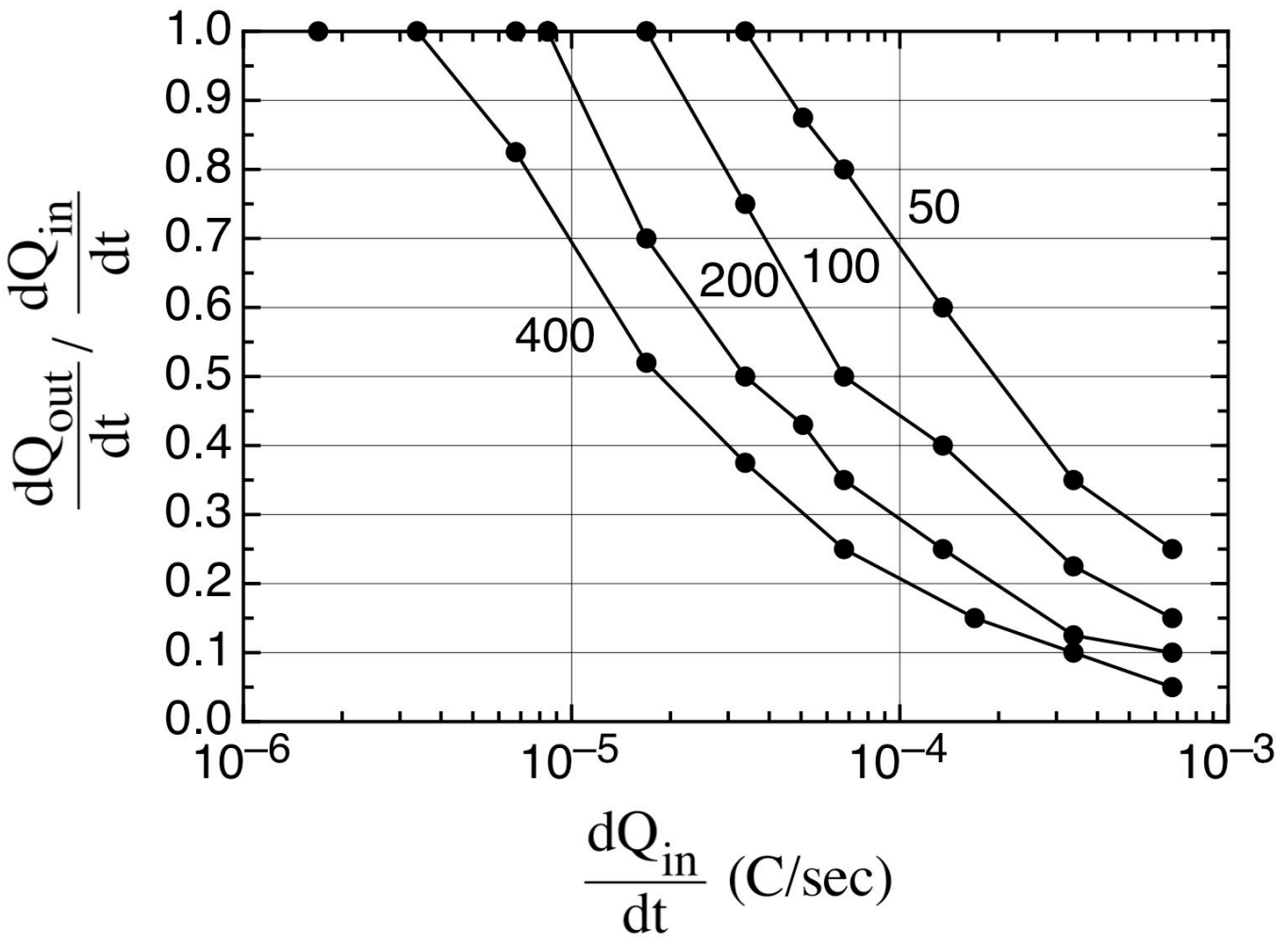

Fig. 7. The fraction of extracted charge, i.e., efficiency of $e^{-}/He^{+}$ extraction, as a function of incident beam rate for different values of buffer gas pressure (in mbar), $E_o$ = -10 V/cm.

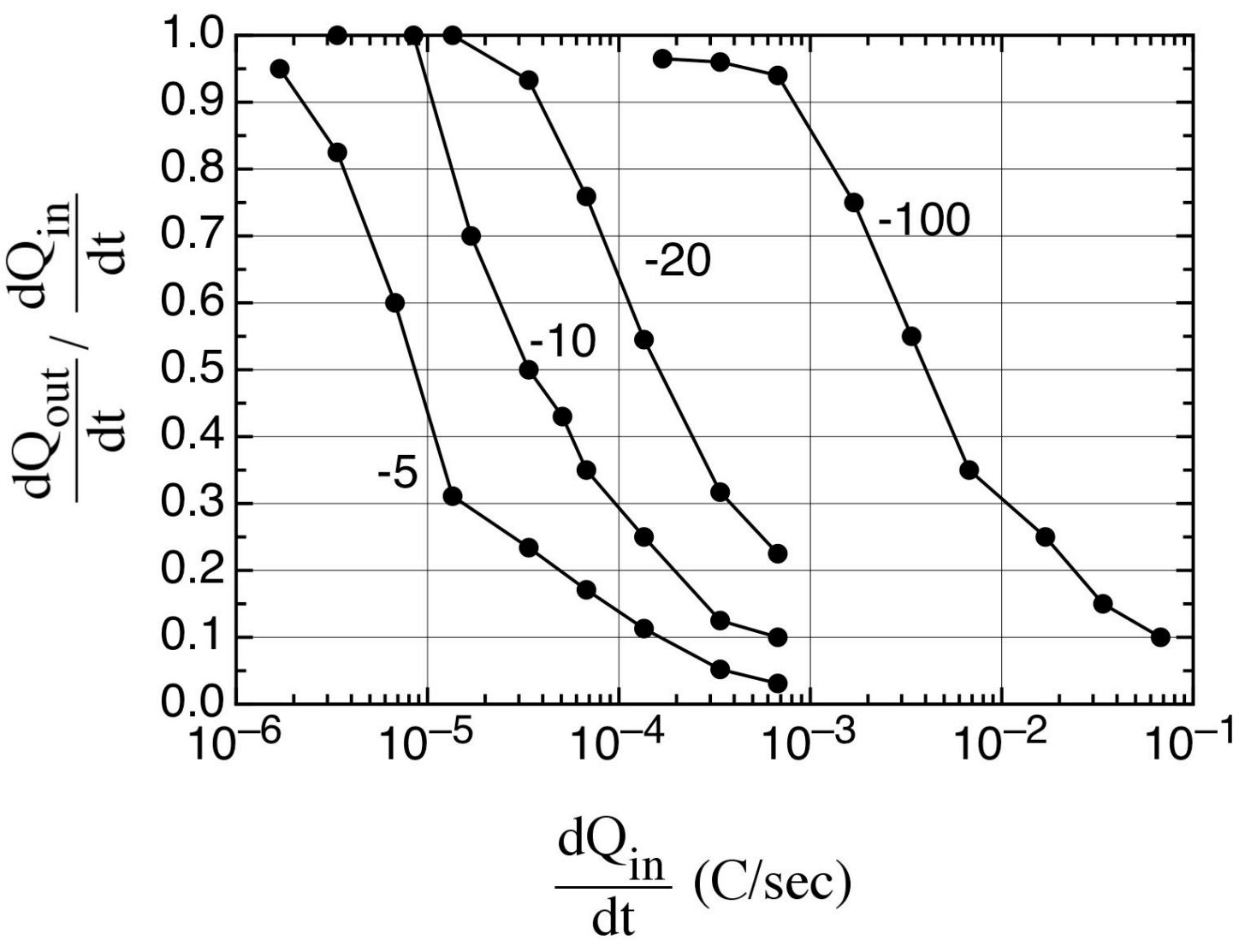

Fig. 8. Similar to Fig. 7, the efficiency of $e^{-}/He^{+}$ extraction as a function of incident beam rate for different values of the applied field $E_o$ (in V/cm), $P$ = 200 mbar.